\documentclass{article}
\usepackage{amsmath,amssymb}
\usepackage{graphicx}
\begin{document}
\title {Effect of electrons on equation of state of porous materials  \\}
\author {Bishnupriya Nayak \\ 
Theoretical Physics Division, Bhabha Atomic Research Centre\\ Mumbai-400 085, India\\}
\date{}
\maketitle

\begin{abstract}
A new equation of state (EOS) is developed for porous materials in which contribution of electrons is considered explicitly. This EOS describes anomalous behaviour of hugoniot of porous substances as observed experimentally. Using this EOS, hugoniot of copper and aluminium are evaluated for different porosities and the agreement with experimental data is good. The present EOS is valid over a wide range of porosities (1 to 10). The contribution of electrons is significant for porosity $\geq 2$. Also, shock and particle velocity curves obtained using this EOS agree well with experimental data.
\end{abstract}

\section{Introduction}
Investigation on dynamic behaviour of porous materials is a topic of current interest due to their shock isolation and attenuation properties. Porous materials are characterized by a factor known as porosity ($\alpha$) which is defined as the ratio of density of normal material to that of porous material. Under high compression porous solids (when $\alpha\geq 2$) show anomalous behaviour i.e. with increasing applied pressure the volume increases instead of decreasing. This behaviour is reflected as turning in the hugoniot observed at high porosity. The underlying physics of anomalous behavior in porous solids is as follows. The presence of pores gives additional contribution to specific internal energy (in form of surface energy of pores). In other words, the associated specific internal energy is high as compared to normal solid. As a consequence when a porous material is compressed (size and number of pores are reduced), temperature (T) of the material increases appreciably. This leads to increase in volume in the initial phase. This behavior is in contrast with that of normal solid where volume decreases with applied pressure. When all the pores are collapsed it behaves like a normal solid. If initial porosity is low ($\alpha < 2$), the number of pores are less and so is the surface energy. As a result the porous material does not expand initially and behaves like normal solid. This peculiar (anomalous) behaviour  was first observed experimentally by Krupnikov \cite{Krupnikov} and Kormer \cite{Kormer}. On this basis Zeldovich and Raizer \cite{Zel2} had given a qualitative picture on shock compression of porous materials. Many other theoretical and experimental studies have been done to predict several models for porous EOS \cite{Herrmann, Holt1, Simons, Trunin, Oh, Dijken, Wu, Viljoen, Menikoff, Huayun}. But most of the models could not explain the anomalous behaviour of hugoniot. It is argued by some authors \cite{Wu, Viljoen} that hugoniot of porous substances obtained from Mie-Gruneisen (MG) EOS \cite{Eliezer} does not reveal the turning of hugoniot as observed experimentally. In 1996 Wu-Jing (WJ) proposed an EOS along isobaric path to describe hugoniot  of porous solids \cite{Wu}. This EOS describes relationship between specific volume and specific enthalpy of a substance through material parameter R which is a function of pressure. This parameter R is  analogous to Gruneisen coefficient $\Gamma$ (function of volume) in MG EOS. \\

Wu-Jing  and Viljoen calculated the parameter R from hugoniot of normal solid without accounting for  electronic contribution. The WJ model was improved by Huayun \emph {etal} by incorporating contribution of electrons to EOS in low temperature regime \cite{Huayun}. In this paper, thermal contribution of electrons is explicitly accounted for to obtain hugoniot of normal solid. During shock compression solid may undergo many transformations (solid-liquid transition, liquid-gas transition, dissociation, ionization etc.). The ionic specific heat of normal solid at high temperature tends to $3 \bar{R}$ where $\bar{R}$ is the universal gas constant per unit mass. After shock compression if the substance is in gaseous phase its specific heat ($C_v$) is $3\bar{R}/2$. These effects have been accounted via temperature and density variations of ionic specific heat and Gruneisen coefficient \cite{Kormer, Grover1971, Trunin1998, Gordeev2008}. Also at very high temperature ($\sim 50,000\,$ K) electrons behave as an ideal gas. An interpolation formula given by Kormer \emph{etal} \cite{Kormer} is used for electronic contribution to EOS. At low T this formula gives the correct limiting behaviour i.e. electronic specific heat is proportional to T and at very high T it tends to ideal gas limit. \\  
 
The paper is organized as follows. In section 2 Rankine-Hugoniot relations across shock front and EOS of material are mentioned. Section 3 describes Wu-Jing and Viljoen methods for EOS of porous substances. Section 4 contains the evaluation of EOS of porous materials. Results and discussion are given in section 5. Finally, the conclusion is given in section 6.    

\section {Shock hugoniot relations and EOS of solids}
When a steady shock front \cite{Zel1} propagates through a material at rest (i.e. initial velocity $U_0$ of material is zero) it compresses the material behind it. Assuming thermodynamic equilibrium in the material ahead and behind the shock front; mass, momentum and energy conservation equations take the following form:
\begin{eqnarray}
\rho_0 U_s = \rho (U_s - U_p) \label{shock_mass}  \\ 
P_0 + \rho_0 U_s^2 = P + \rho {(U_s - U_p)}^2 \label{shock_mom}  \\
E_0 + \frac{P_0}{\rho_0} +  \frac{1}{2}U_s^2 = E + \frac{P}{\rho} + \frac{1}{2}{(U_s - U_p)}^2  \label{shock_ener}
\end{eqnarray}
where $\rho_0$, $P_0$, $E_0$ are respectively, density, pressure and specific internal energy of the substance ahead of the shock front at ambient condition. $\rho$, $P$, $E$ are the same flow variables behind the  front.  $U_s$ is the Shock velocity and $U_p$ is the material velocity behind the front. Generally it is observed from experimental shock hugoniot data that a linear relationship holds between $U_s$ and $U_p$ \cite{Meyer} i.e.
\begin{equation}
U_s = c_0 + s_1 U_p	\label{us-up}
\end{equation}
where $c_0$ is sound velocity at initial density $\rho_0$ and $s_1$ is empirical constant. The values of $c_0$ and $s$ are tabulated in ref.\cite{Meyer} for different materials. All the three conservation equations (Eq(\ref{shock_mass}-\ref{shock_ener})) are known as Rankine-Hugoniot relations. Substituting Eq(\ref{shock_mass}) and Eq(\ref{shock_mom}) in Eq(\ref{shock_ener}) one can obtain hugoniot relation across the shock front which is expressed as follows:
\begin{equation}
E - E_0 = \frac{1}{2}(P + P_0)(v_0 - v)	\label{hugorel}
\end{equation}
where $v_0$($=1/\rho_0$) and $v$($=1/\rho$) are the specific volumes before and after shock compression. Eq(1-4) have  five unknown quantities i.e. $\rho$, $P$, $E$, $U_s$, $U_p$. In order to evaluate these flow variables uniquely we need equation of state(EOS) of material. EOS is the thermodynamic relationship among the flow variables like $v$, $\rho$, $P$, $E$ \emph{etc.} \\

Most commonly used EOS is the Mie-Gruneisen EOS \cite{Eliezer}. It relates a state (P,V,E) to the pressure and specific internal energy of a reference state at the same specific volume\cite{Meyer}. Mathematically, it can be expressed as:
\begin{equation}
P - P_{ref} = \frac{\Gamma}{v} (E - E_{ref})	\label{Mie}
\end{equation}
where $\Gamma$ is the Gruneisen-coefficient. $P_{ref}$ and $E_{ref}$ are the pressure and specific internal energy of the reference state. The reference state may be zero-kelvin isotherm or hugoniot state of a substance. In general, total specific internal energy  and pressure of a substance have three components. These are: (i) cold or elastic  (ii) ionic and (iii) electronic. The cold component arises due to inter-atomic bonding and zero point vibrational energy. So, it depends only on volume. The atomic contribution comes because of phonon vibration and the electronic contribution arises due to thermal excitation of electrons. Mathematically one can write:
\begin{eqnarray}
E(v,T) = E_c (v) + E_{Ta}(v,T) + E_{Te}(v,T) \label{geneosE} \\
P(v,T) = P_c (v)+ P_{Ta}(v,T) + P_{Te}(v,T) \label{geneosP} 
\end{eqnarray}
where $E_c$, $P_c$; $E_{Ta}$, $P_{Ta}$; $E_{Te}$, $P_{Te}$ are the cold or elastic, atomic and electronic components of specific internal energy and pressure respectively. \\

\section{Wu-Jing and Viljoen method for porous materials}
In 1957 Rice and Walsh \cite{Rice} proposed an EOS for water in terms of specific enthalpy $H$ and specific volume $v$ in the pressure range 25 to 250 kilobar. This EOS was an empirical fit to experimental data. Wu and Jing derived the same EOS from thermodynamic considerations along isobaric path with assumption that specific heat at constant pressure ($C_p$) remains same. According to their formula the EOS can be written as:
\begin{equation}
v - v_c = \frac{R}{P} (H - H_c)	\label{WJeos}
\end{equation} 
where $v_c$, $H_c$ are the specific volume and specific enthalpy on zero-kelvin isotherm ($P_c$), P is the pressure and R is the material parameter which is a function of pressure. The detailed description of the parameter R is given in ref. \cite{Wu}. By analogy to MG EOS, this EOS relates the state (v,P,H) to specific volume and specific enthalpy on zero-kelvin isotherm at the same pressure. Eq(\ref{WJeos}) is applicable to both normal and porous solids. For porous materials, the EOS can be written as:
\begin{equation}
v_h' - v_c' = \frac{R}{P} (H_h' - H_c')	\label{WJporeos}
\end{equation} 
where prime refers to thermodynamic  quantities of porous material and subscript $h$ stands for hugoniot state. The parameter R remains same for normal and porous substances as $R=R(P)$. The specific enthalpies on zero-kelvin isotherm and on hugoniot are given as:
\begin{eqnarray}
H_c'= P v_c' +E_c' \\
H_h'=E_{00}+\frac{1}{2}P_1 (v_{00}-v_1)+\frac{1}{2}P (v_1+v_h')  \label{WJentha}
\end{eqnarray}
where $E_{00}$, $v_{00}$ are the initial specific internal energy and initial specific volume of porous solid. The subscript 1 stands for the hugoniot elastic limit (HEL) of porous material. Wu and Jing used an approximate model  proposed by Carroll and Holt \cite{Holt2} to predict HEL of porous solids. The calculation of HEL requires cold pressure($P_c$), initial porosity($\alpha$) and material yield strength(Y). In WJ method, many parameters like constants associated with $P_c$, $Y$, HEL, isoentropic bulk modulus($k_s$) are needed to evaluate EOS of porous substances. \\
Therefore Viljoen \cite{Viljoen} modified the WJ method without considering $Y$, HEL and $k_s$. Viljoen method uses cold ($P_c$) and hugoniot pressure ($P_h$) of normal solid to evaluate EOS of porous materials. The pressure hugoniot is obtained using MG  EOS and the form is:
\begin{equation}
P_h = \frac{P_c - \Gamma \rho E_c}{1- \frac{\Gamma}{2} \left( \frac{v_0}{v_h} -1\right)}	\label{MGeos}
\end{equation} 
where $v_h$ is specific volume on hugoniot of normal material. Eq(\ref{MGeos}) is used to calculate specific  volume on hugoniot under isobaric condition but it does not account for  excitation of electrons. As mentioned in section 2, the pressure hugoniot contains all the three components i.e. cold, ionic and electronic. In next section the evaluation of EOS of porous materials with electronic contribution is described.\\

\section{EOS of porous materials} 
The total specific internal energy and pressure have cold or elastic, ionic and electronic components as given in earlier section. The components used to evaluate EOS of porous solids are described below. 

\subsection{Cold component}
\begin{figure}
\centering
\includegraphics[width=1\textwidth]{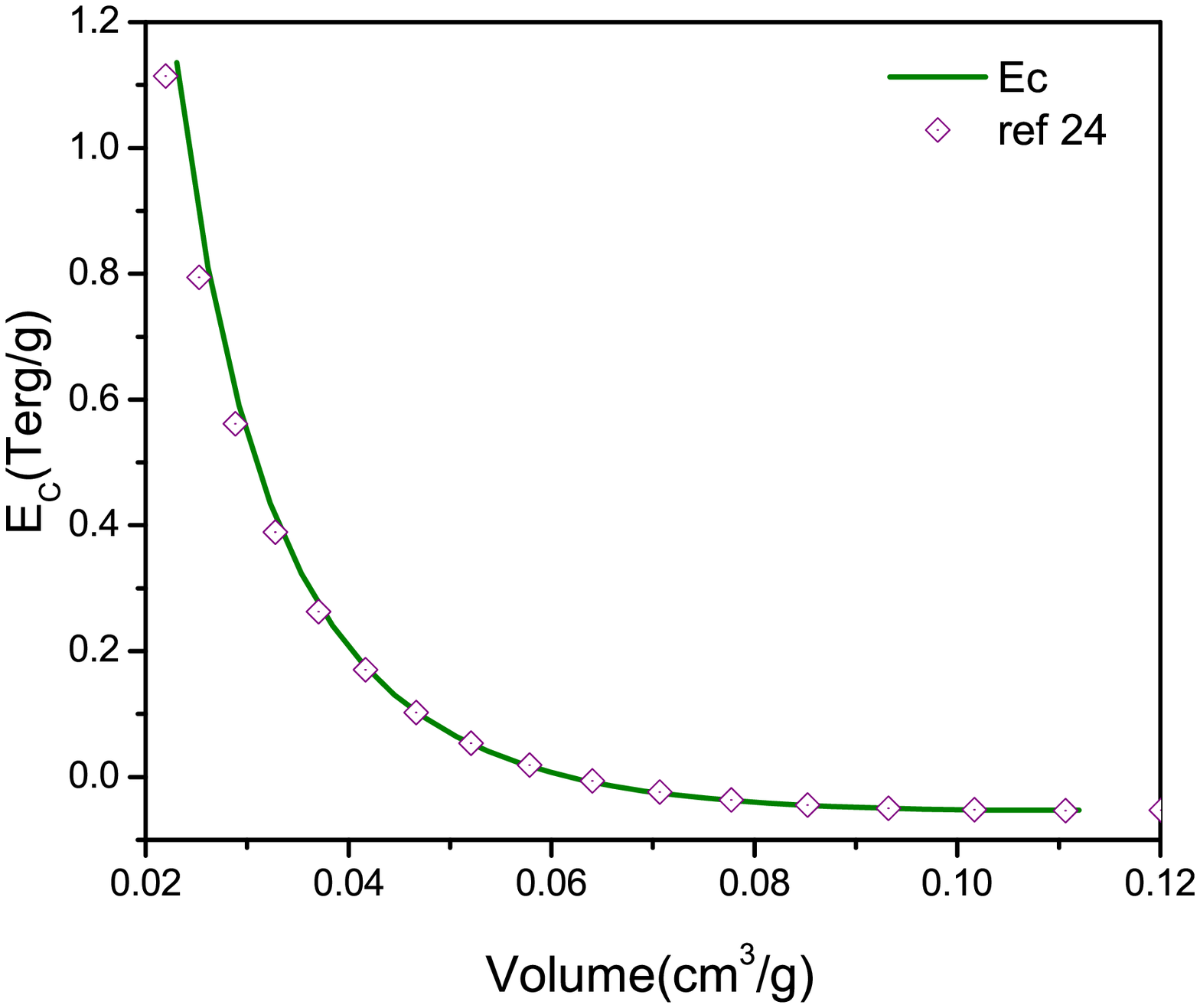}
\caption{Comparison of $E_c$ with \emph{ab initio} calculation for Cu.}
\label{Cu-cold}
\end{figure}

\begin{figure}
\centering
\includegraphics[width=1\textwidth]{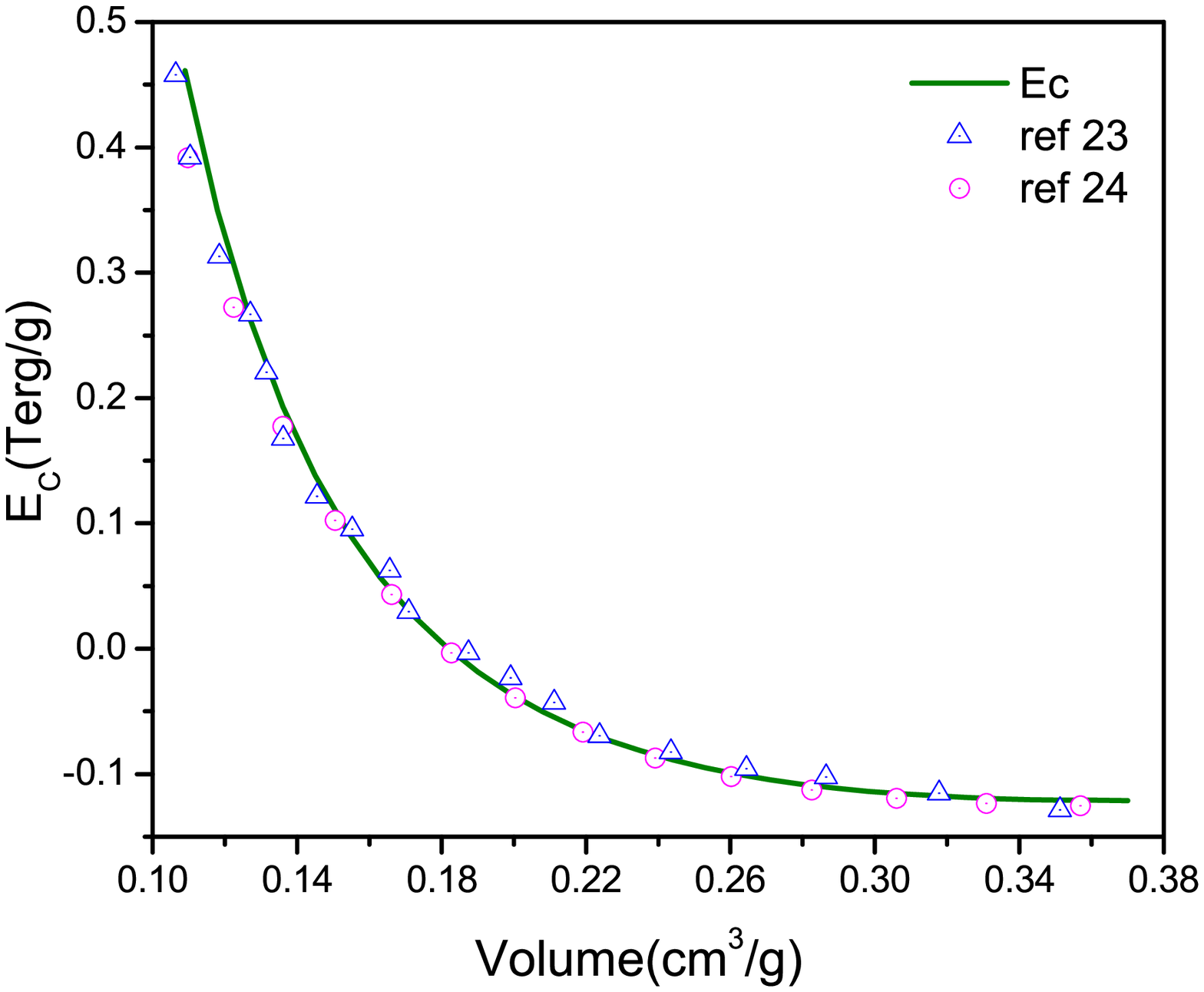}
\caption{Comparison of $E_c$ with \emph{ab initio} calculations for Al.}
\label{Al-cold}
\end{figure}

 The cold or elastic pressure $P_c$ is given as \cite{Guskov}:
\begin{equation}
 P_c= \frac {\rho_0 c_0^2}{n-m} \left[\left( \frac{v_0}{v_c}\right)^n-\left( \frac{v_0}{v_c}\right)^m \right] \label{Pc}
\end{equation} 
where n and m are fitting constants. The constants are chosen so that they satisfy the following condition.
\begin{equation}
n+m = B_0' 
\end{equation} 
where $B_0'$ is the pressure derivative of bulk modulus at ambient condition. The elastic specific internal energy $E_c$ can be obtained by integrating $P_c$ over the volume with initial condition $E_c(v_0)=0$ and its form is:
\begin{equation}
 E_c= \frac {c_0^2}{(n-m)(n-1)} \left[\left( \frac{v_0}{v_c}\right)^{n-1}-\frac{n-1}{m-1}\left( \frac{v_0}{v_c}\right)^{m-1} \right]+\frac {c_0^2}{(n-1)(m-1)} \label{Ec}
\end{equation} 
The cold component of specific internal energy matches quite well with data obtained from \emph{ab initio} calculations \cite{Wang,Sai} as shown in fig \ref{Cu-cold} and \ref{Al-cold}.\\

\subsection{Ionic component}
When a material is compressed by shock wave its temperature increases and it may undergo various transformations (solid-liquid transition, liquid-gas transition, dissociation, ionization etc). After shock compression if the material is in gaseous phase its specific heat becomes $3\bar{R}/2$. But for a solid when $T$ is greater than Debye temperature the specific heat tends to $3\bar{R}$. For most of solids, the Debye temperature lies between $300$ K to $500$ K. Similarly when $T\rightarrow \infty$, the Gruneisen-coefficient attains a limiting value $2/3$ as in ideal gas. The change in temperature also affects density of substance. Therefore, temperature and density variations of specific heat and Gruneisen coefficient are accounted for in the ionic contribution \cite{Kormer, Grover1971, Trunin1998, Gordeev2008}. The formulas for specific heat and effective Gruneisen coefficient ($\lambda$) are:
\begin{eqnarray}
C_{v}= \frac{3}{2} \bar{R}  \left [1+\frac{g^2(\rho)}{\{g(\rho) +T\}^2} \right]  \label{modCv} \\
\lambda = \frac{2}{3}\left[ \frac{3 \Gamma g(\rho)+ T}{2 g(\rho) +T} \right]  \label{modgam}
\end{eqnarray} 
where $g(\rho)$ is a parameter that depends on density. The parameter is calculated from Lindemann's formula as mentioned in ref. \cite{Grover1971} with assumption $\Gamma \rho =\Gamma_0 \rho_0$, where $\Gamma_0$ is the Gruneisen coefficient of normal material at ambient condition. The form of $g(\rho)$ is:
\begin{equation}
g(\rho) = g_0 \left(\frac{v}{v_0}\right)^{2/3} \mbox{exp}\left[2 \Gamma_0 \left(1-\frac{v}{v_0}\right)\right] \,. \label{gr}
\end{equation}
The constant $g_0$ is obtained from the relation $g_0 = Q_{bond}/(C_{v}/2)$ where $Q_{bond}$ represents enthalpy of vaporization. Using Eq(\ref{modCv}) and (\ref{modgam}), one can obtain ionic specific internal energy and pressure. The expressions for ionic components are:
\begin{eqnarray}
E_{Ta}= \left \{\frac{2 g(\rho)+ T}{g(\rho) +T} \right\} \frac{3}{2} \bar{R} T \label{ET}\\
P_{Ta}= \left \{\frac{3 \Gamma g(\rho)+ T}{g(\rho) +T} \right\} \rho \bar{R} T \,. \label{PT} 
\end{eqnarray}
From Eq(\ref{modCv}),(\ref{modgam}),(\ref{ET}) and (\ref{PT}) it is clear that when $T\rightarrow 0$
\begin{equation}
C_v \approx  3\bar{R} \; ; \; \lambda \approx \Gamma \; ; \;P_{Ta} \approx \Gamma \rho E_{Ta}	\, :\mbox{Mie-Gruneisen EOS}  \label{atlowlt}
\end{equation}
and when $T\rightarrow \infty$
\begin{equation}
C_v \approx \frac{3\bar{R}}{2} \; ; \; \lambda \approx \frac{2}{3}\; ; \;P_{Ta} \approx \rho  \bar{R} T	\,: \mbox{Ideal gas EOS}\,.  \label{athighlt}
\end{equation}
The formulas for ionic components are valid in a wide range of density and temperature.   

\subsection{Electronic component}
An interpolation formula proposed by Kormer \emph{etal} \cite{Kormer} for thermal energy of electrons is: 
\begin{equation}
E_{Te}=\frac{b^2}{\beta} \mbox{ln} cosh\left(\frac{\beta T}{b}\right) \label{Ee} \\
\end{equation}
where $b=1.5 Z \bar{R}$ and $Z$ is the atomic number of material. $\beta$ is the coefficient of electronic specific heat. The above fitting formula is obtained from Latter's data \cite{Latter} who had calculated thermal energy of electrons using Thomas-Fermi equation for $T\neq 0$. The corresponding thermal pressure for electrons is
\begin{equation}
P_{Te}=\Gamma_e \rho \frac{b^2}{\beta} \mbox{ln} cosh\left(\frac{\beta T}{b}\right) \label{Pe}
\end{equation}
where 
\begin{equation}
\Gamma_e = -\frac{d \mbox{ln} \beta}{d \mbox{ln} \rho} \, . \label{ge}
\end{equation}
From Eq(\ref{Ee}) it is obvious that when $T\rightarrow 0$ 
\begin{equation}
E_{Te}= \frac{1}{2} \beta T^2	\label{elowlt}
\end{equation}
and when $T\rightarrow \infty$
\begin{equation}
E_{Te}= \frac{3}{2} Z \bar{R}T \, . \label{ehighlt}
\end{equation}
Like ions, the electronic components are also valid in a wide range of density and temperature.

\subsection{EOS along isobaric path}
For a solid material one can write EOS along isobaric path as:
\begin{equation}
v_h(P,T) - v_c(P) = \frac{R(P)}{P} \{H_h(P,T) - H_c(P)\}	\label{WJsolideos}
\end{equation} 
where the enthalpies on zero-kelvin isotherm and hugoniot state are:
\begin{eqnarray}
H_c(P) = P v_c + E_c(v_c) \\
H_h(P,T) = E_h(v_h,T) + P v_h   \,.  \label{norenthalpy}
\end{eqnarray}
After substituting $H_c$ and $H_h$ in Eq(\ref{WJsolideos}), it becomes
\begin{equation}
v_h - v_c = \frac{R}{P} \left \{ E_{0} + \frac{1}{2}P_0 (v_{0}-v_h) + \frac{1}{2}P (v_0+v_h)- E_c - P v_c\right\}    \label{R1} 
\end{equation} 
Along isobaric path the pressure on zero-kelvin isotherm is same as the pressure on hugoniot of normal and porous solids i.e. $P_c=P_h=P_{por} = P$. To determine the parameter R from Eq(\ref{R1}) one needs to know volume on hugoniot of normal solid. The $v_h$ can be calculated along isobaric path by solving two nonlinear equations simultaneously using Newton-Raphson method. The nonlinear equations are
\begin{eqnarray}
P = P_c (v_h)+ P_{Ta}(v_h,T_h) + P_{Te}(v_h,T_h) \label{totalP} \\
E_{h}-E_0= \frac{1}{2}(P + P_0)(v_0-v_h)   \label{hugo_relt}
\end{eqnarray}  
where 
\begin{equation}
E_h = E_c (v_h)+ E_{Ta}(v_h,T_h) + E_{Te}(v_h,T_h). \label{totalE} 
\end{equation} 
The calculation of each ($v_h,T_h$) on hugoniot requires an initial guess ($v_{ig},T_{ig}$). The guess values are obtained from hugoniot relation of normal solid under isochoric condition using bisection method. Now the parameter R can be determined from the following equation.
\begin{equation}
R = \frac{P_c (v_h - v_c)} {\left \{ E_{0} + \frac{1}{2}P_0 (v_{0}-v_h) + \frac{1}{2}P_c (v_0+v_h)- E_c - P_c v_c\right\}} \label{Rp} 
\end{equation}
The same R is used to evaluate EOS of porous substances and the EOS is:
\begin{equation}
v_{por} - v_c = \frac{R}{P_c} \left \{ E_{00} + \frac{1}{2}P_{00} (v_{00}-v_{por}) + \frac{1}{2}P_c (v_{00}+v_{por})- E_c - P_c v_c\right\}    \label{hugopor} 
\end{equation}
where assumption is that $E_{00}=E_{0}$ and $P_{00}=P_0$. The constants used to evaluate EOS of porous materials are listed in table \ref{tab:consts}. 

\begin{table}
\caption{\emph{Constants used to evaluate EOS of porous materials.}}
\label{tab:consts} 
\begin{quote}
\centering
\begin{tabular}{|c c c |}  
\hline
Constants& Cu& Al \\
\hline
\hline
$\rho_0$(g/cm$^3$)& 	8.93&	2.702	\\
$c_0$(km/sec)&	4.0&	5.2	\\
n&	3&	3 	\\
m&	2& 1.2	\\
$\Gamma_0$&	2.0&	2.1 \\
$R\times 10^{6}$(Terg/(g K))&	1.3094&  3.0794 \\
$Q_{bond}\times 10^{-2}$(Terg/g)&   4.73&  10.87 \\	
$g_0$(K)&  24078.5&  23525.3  \\
$b\times10^{5}$(Terg/(g K))&   5.696&	6.005  \\
$\beta\times10^{12}$(Terg/(g K$^2)$)&   109.34&	500.33  \\
\hline  
\end{tabular}
\end{quote}   
\end{table} 

\section{Results and discussion}
\begin{figure}
\centering
\includegraphics[width=1\textwidth]{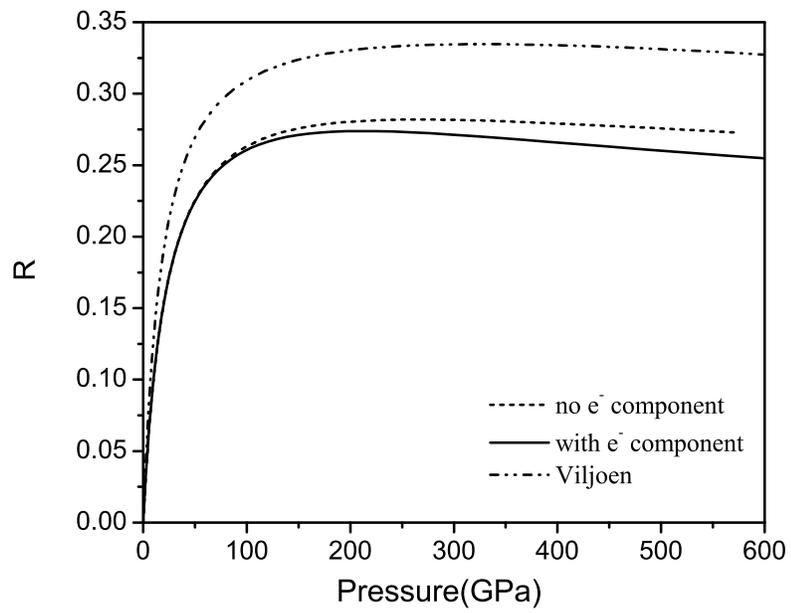}
\caption{R vs P curve for Cu. solid line-- parameter R with electronic contribution; dashed line-- parameter R without electronic contribution; dash dot dot line-- Viljoen method.}
\label{Cu-R}
\end{figure}

\begin{figure}
\centering
\includegraphics[width=1\textwidth]{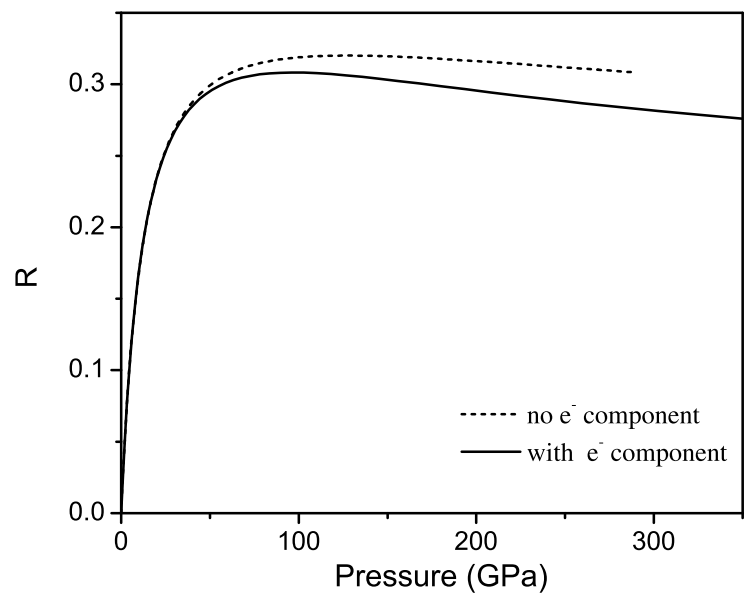}
\caption{R vs P curve for Al. solid line-- parameter R with electronic contribution; dashed line-- parameter R without electronic contribution}
\label{Al-R}
\end{figure}

\begin{figure}
\centering
\includegraphics[width=1\textwidth]{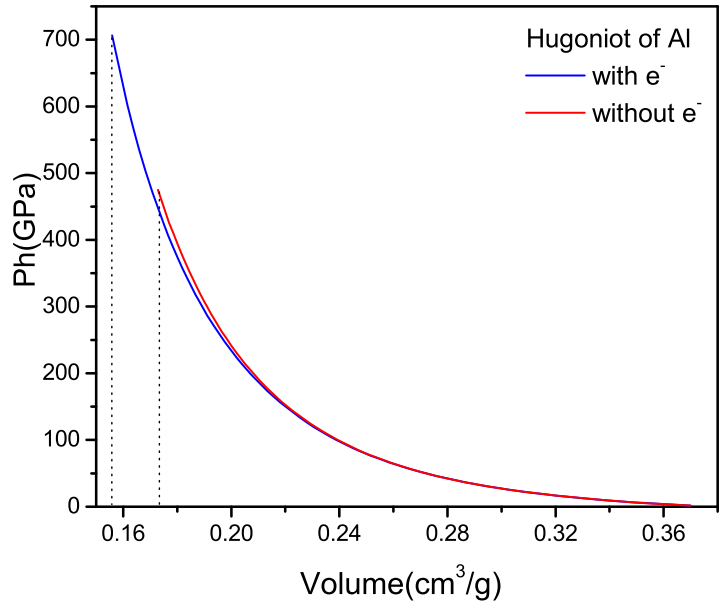}
\caption{For same temperature compression is more due to electrons than without electrons which is shown in dotted lines for Al.}
\label{hugoAl}
\end{figure}

The parameter R is determined for Cu and Al using general EOS as described in the previous section. The R vs P curves are shown in fig \ref{Cu-R} and \ref{Al-R}. We have compared the parameter R obtained from present model with Viljoen model for Cu. It is evident from fig \ref{Cu-R} that the values of R obtained from Viljoen model are higher as compared to the present model (without electronic component). This is primarily due to the difference in models used for cold pressure $P_c$. The $P_c$ used in this work is slightly higher than the $P_c$ ($P_c^{vj}$) used by Viljoen. The difference between $v_h$ and $v_c$ mainly decides the value of R as given in Eq(\ref{Rp}). Since the $P_c$ used in this work lies above that of $P_c^{vj}$ the term $v_h-v_c$ is lower along isobaric path, thus decreasing the value of R. However, inclusion of electronic contribution to present EOS leads to further decrease in the value of R (which is clear from fig \ref{hugoAl}). Hence the electronic contribution to EOS can't be ignored.\\

\begin{figure}
\centering
\includegraphics[width=1\textwidth]{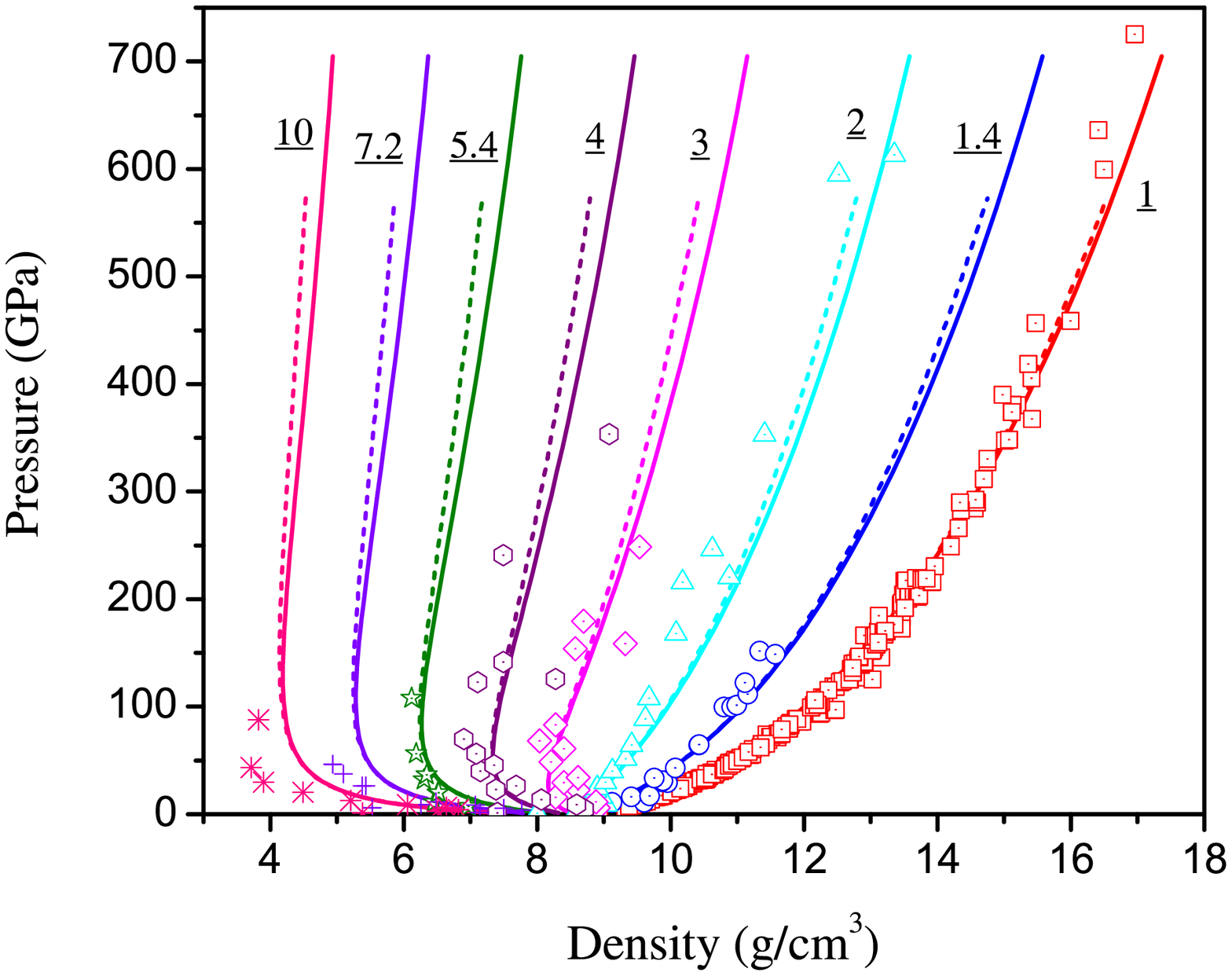}
\caption{Comparison of theoretically predicted hugoniot of porous Cu with experimental data. solid line-- hugoniot with electronic contribution; dashed line-- hugoniot without electronic contribution; symbols-- experimental shock data \cite{shock-data}.}
\label{Cu-por}
\end{figure}

\begin{figure}
\centering
\includegraphics[width=1\textwidth]{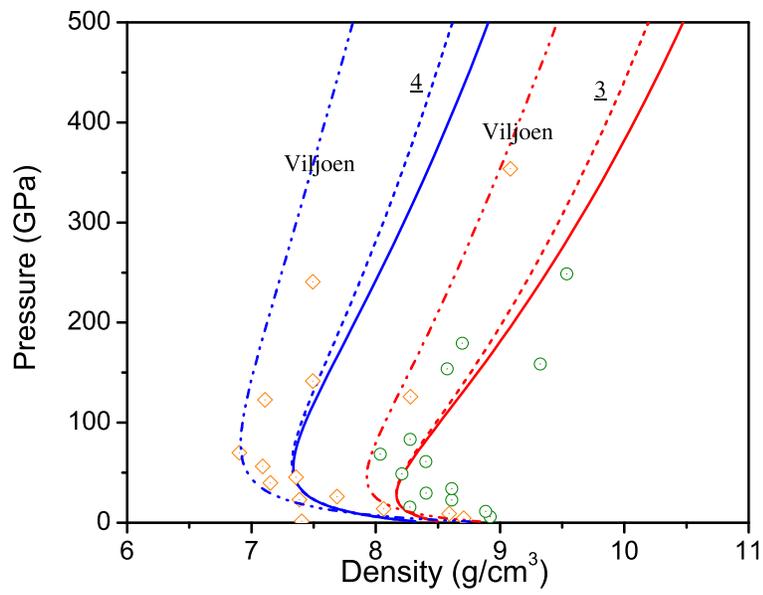}
\caption{Comparison of theoretically predicted hugoniot of Cu with experimental data and Viljoen method for $\alpha =3, 4$. solid line-- hugoniot with electronic contribution; dashed line-- hugoniot without electronic contribution; symbols-- experimental shock data \cite{shock-data}; dash dot dot line-- Viljoen method}
\label{Cu-vj}
\end{figure}

\begin{figure}
\centering
\includegraphics[width=1\textwidth]{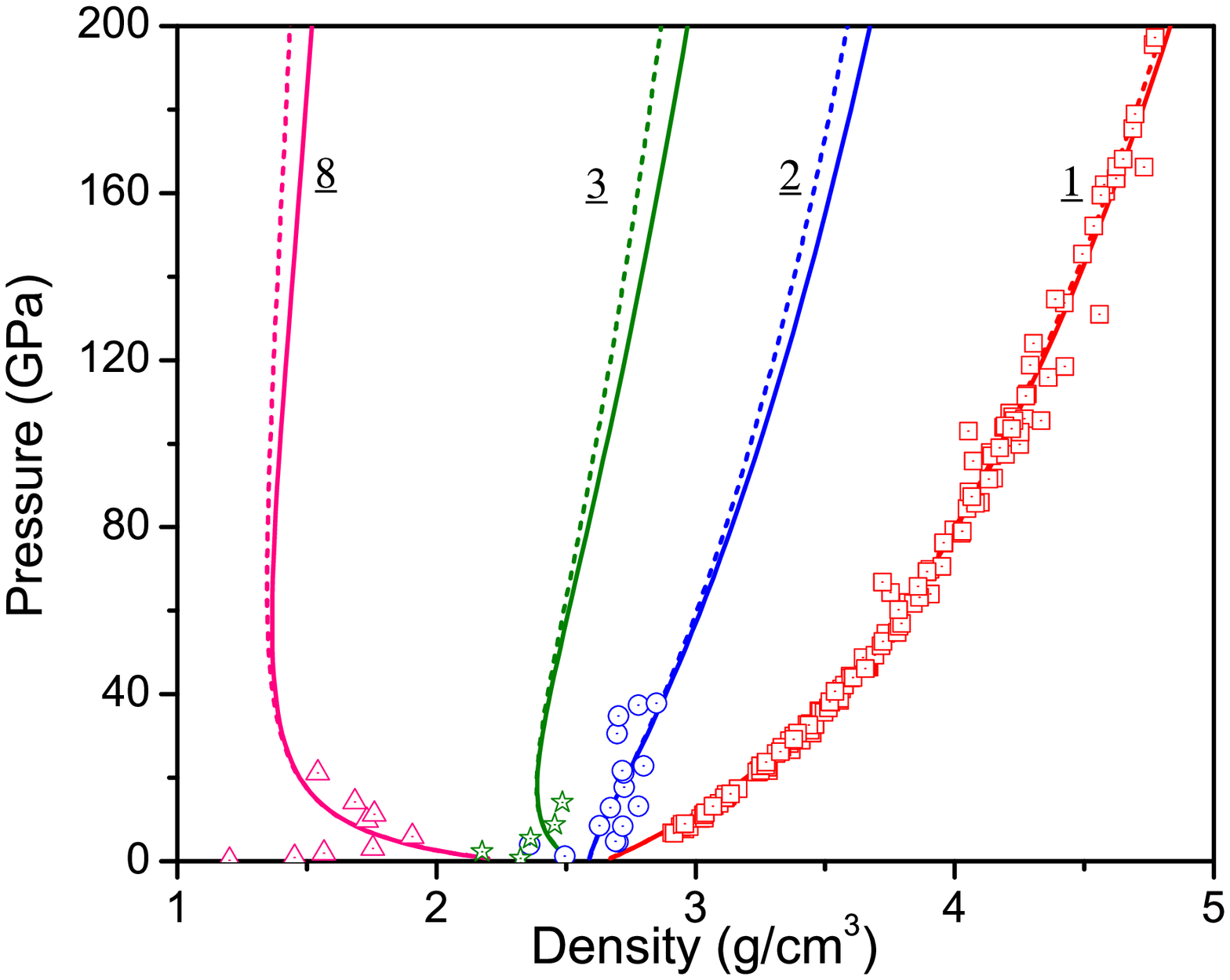}
\caption{Comparison of theoretically predicted hugoniot of porous Al with experimental data. solid line-- hugoniot with electronic contribution; dashed line-- hugoniot without electronic contribution; symbols-- experimental shock data \cite{shock-data}.}
\label{Al-por}
\end{figure}

Hugoniot of porous Cu and Al are obtained using the parameter R. The hugoniot of Cu with and without electrons are shown in fig \ref{Cu-por} for different initial porosities i.e. $\alpha = 1, 1.4, 2, 3, 4, 5.4, 7.2, 10$. When $\alpha\geq 2$, anomalous behavior is observed in hugoniot. Good agreement is observed between theoretically predicted hugoniot and experimental shock data for porous Cu. Fig \ref{Cu-vj} shows the comparison of hugoniot of Cu obtained from present model with Viljoen model for $\alpha =3, 4$. The present model agrees better with experimental data as compared to Viljoen method. In fig \ref{Al-por} the hugoniot of Al with and without electrons is shown for $\alpha = 1, 2, 3, 8$. We find the agreement of present calculation with experimental data is reasonably good. It is clear from fig \ref{Cu-por} and \ref{Al-por} that electronic contribution is significant for $\alpha\geq 2$ and hence can't be ignored in EOS of porous substances.\\

\begin{figure}
\centering
\includegraphics[width=1\textwidth]{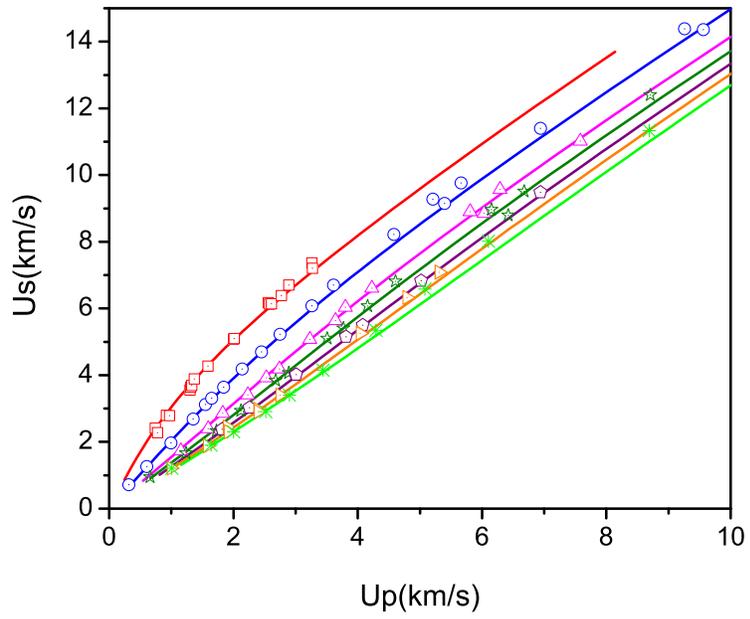}
\caption{Us vs Up curve for porous Cu. red- $\alpha$=1.4; blue- $\alpha$=2; magenta- $\alpha$=3; olive- $\alpha$=4; purple- $\alpha$=5.4; orange- $\alpha$=7.2; green- $\alpha$=10; symbols-- experimental shock data \cite{shock-data}.}
\label{Cu-Up}
\end{figure}

\begin{figure}
\centering
\includegraphics[width=1\textwidth]{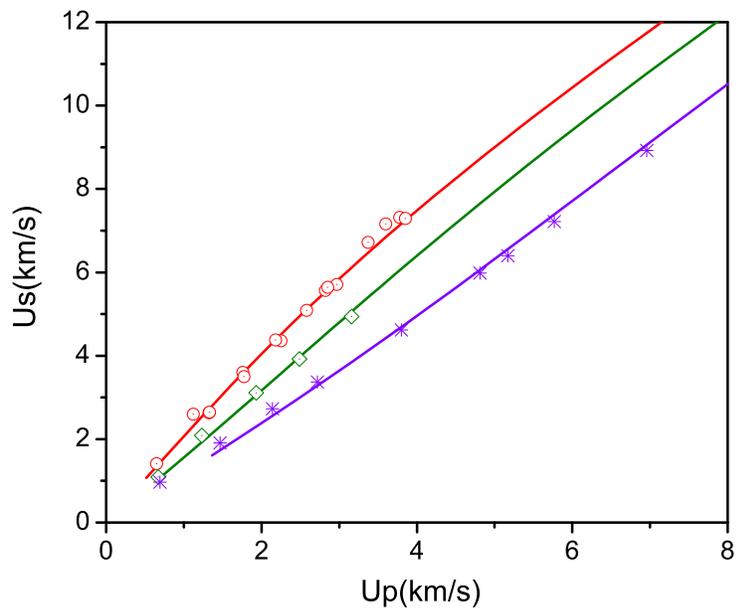}
\caption{Us vs Up curve for porous Al. red- $\alpha$=2; olive- $\alpha$=3; violet- $\alpha$=8;  symbols-- experimental shock data \cite{shock-data}.}
\label{Al-Up}
\end{figure}

The shock and particle velocity curves obtained from present EOS model are shown in fig \ref{Cu-Up} and \ref{Al-Up} for porous Cu and Al. It is evident from fig \ref{Cu-Up} and \ref{Al-Up} that linear relationship between $U_s$ and $U_p$ does not hold for porous materials as given in Eq(\ref{us-up}). The $U_s$ and $U_p$ curves of these substances can be fitted with higher order polynomials for different empirical constants $s_1$, $s_2$, $s_3$ etc. The form of polynomial is: 
\begin{equation}
U_s = c_0 + s_1 U_p + s_2 U_p^2 + s_3 U_p^3 +...	\label{usup-por}
\end{equation}
The theoretically predicted $U_s-U_p$ curves agree well with experimental data for all porosities of Cu and Al. Hence it is clear that the present EOS model is valid for high porosities. \\


\section{Conclusion} 
We have presented an EOS for porous materials including ionic and electronic contributions explicitly. According to our knowledge this is the first time that the contributions have been considered explicitly. Earlier works were based on MG EOS. The low and high temperature limits of ionic and electronic specific heat are included in the present model. This EOS is valid over a wide range of temperatures and densities i.e. from solid to gas phase. Theoretically predicted hugoniot of porous Cu and Al using this present model agrees well with the experimental data. The significance of electronic contribution to EOS is reflected in the value of R as well as on hugoniot. The $U_s-U_p$ curves obtained from the present EOS model for porous Cu and Al agree well with experimental data and they reveal the new EOS model is valid for high porosities. 
 
\subsection*{Acknowledgment}
I am grateful to Dr. S.V.G. Menon  who suggested this area of research. I am thankful to Chandrani Bhattacharya and Madhusmita Das for useful discussions. Also I am thankful to Head, ThPD, BARC for his kind support to this present work.

\newpage

\newpage

\newpage

\newpage

\newpage

\newpage

\newpage

\newpage

\newpage

\newpage
\end{document}